\documentclass[12pt,a4paper]{article}

\usepackage{latexsym}

\begin{document}


\title{ Time dependence of moments of an exactly solvable Verhulst model under random perturbations
\thanks{This
paper was written with partial financial support from the RFBR
grant 06-01-00814. }}

\author{ V.M.~Loginov}

   \maketitle
\begin{center}
Center of Interdisciplinary Researches of Krasnoyarsk State
Pedagogical University\\
 ul. Lebedevoi, 89.
Krasnoyarsk  660049 Russia\\[1ex]

e-mail: \\
\texttt{loginov@imfi.kspu.ru}\\

\end{center}

\medskip

\begin{abstract}

 Explicit expressions for one point  moments corresponding to stochastic Verhulst model driven by Markovian
coloured dichotomous noise are presented. It is shown that the
moments are the given functions of a decreasing exponent. The
asymptotic behavior (for large time) of the moments is described by
a single decreasing exponent.

\end{abstract}

\section{Introduction}\label{sec 1}
There are a lot of papers devoted to description of the temporary
evolution of moments with exactly solvable nonlinear stochastic
equations. In \cite{glts2006} we gave some general procedure to
explicitly solve the master equations of hyperbolic type
corresponding to nonlinear stochastic equations driven by
dichotomous noise. The method is based on a generalization of
Laplace factorization method \cite{ts05,ts06}. As an example we have
considered complete exact nonstationary solution of the master
equations for probability distribution corresponding to stochastic
Verhulst
 model.

In this paper we calculate one points moments of arbitrary degree
and discuss its time evolution. Let us consider nonlinear
stochastic dynamical system
\begin{equation}\label{Ver}
\dot{x} = p(x)+\alpha(t) q(x),
\end{equation}
where $x(t)$ is the dynamical variable, $p(x)$, $q(x)$ are given
functions of $x$, $\alpha(t)$ is the random function with known
statistical characteristics. The model (\ref{Ver}) arises in
different applications (see for example \cite{HL84,Kampen} and
bibliography therein). An important application of this model
consists in study of noise-induced transitions in physics,
chemistry and biology. The functions $p(x)$, $q(x)$ are often
taken polynomial. For example, if we set $p(x)=p_1x+p_2x^2$,
$q(x)=q_2x^2$, $p_1>0$, $p_2<0$, $|p_2|>q_2>0$, then the equation
(\ref{Ver}) describes the population dynamics when resources
(nutrition) fluctuate (Verhulst model).
 In the following we will assume $\alpha(t)$ to be
binary (dichotomic) noise $\alpha(t)=\pm 1$ with switching frequency
$2\nu>0$. As one can show (see \cite{ShL78}), the averages
$W(x,t)=\langle \widetilde W(x,t)\rangle$ and $W_1(x,t)=\langle
\alpha(t)\widetilde W(x,t)\rangle$ for the probability density
$\widetilde W(x,t)$ in the space of possible trajectories $x(t)$ of
the dynamical system (\ref{Ver}) satisfy a system  (also called
``master equations"):
\begin{equation}\label{WW}
   \left\lbrace  \begin{array}{l}
 W_t + \left(p(x)W\right)_x  + \left(q(x)W_1\right)_x =0, \\[0.5em]
 (W_1)_t + 2\nu W_1 +\left(p(x)W_1\right)_x  + \left(q(x)W\right)_x
 =0.
 \end{array}\right.
\end{equation}
We suppose that the initial condition $W(x,0)=W_0(x)$ for the
probability distribution is nonrandom. This implies that the
initial condition for $W_1(x,t)$ at $t=0$ is zero: $W_1(x,0)=
\langle \alpha(0)\widetilde W(x,0)\rangle=\langle \alpha(0)\rangle
W_0(x)=0$. The probability distribution $W(x,t)$ should be
nonnegative and normalized for all $t$: $W(x,t)\geq 0$,
$\int_{-\infty}^{\infty}W(x,t)\, dx \equiv 1$.

In \cite{glts2006} we have obtained the following explicit form of
the complete solution of the system (\ref{WW}) for probability
distribution $W(x,t)$:
\begin{equation}\label{3}
\begin{array}{l} W(x,\tau)=
\frac{1}{2}e^{-\tau}\left\{\delta\left(x-\frac{e^{-\tau}x_*}{1+(p_2
 +q_2)(e^{\tau}-1)x_*}\right) +
 \delta\left(x-\frac{e^{-\tau}x_*}{1-(p_2
 -q_2)(e^{\tau}-1)x_*}\right)\right\}+{}\\[1em]
 \frac{1}{2q_{2}x^{2}}\left\{ H \left(
\frac{x}{e^{\tau}(1+(p_2
 -q_2)-x(p_2 -q_2))}-x_*\right) - H\left(\frac{x}{e^{\tau}(1+(p_2
 +q_2)-x(p_2 + q_2))}-x_*\right)\right\},
\end{array}
\end{equation}
where $\tau=\nu t$ is the dimensionless time and $x_*$ is an initial
value for (\ref{Ver}),
$H(z)=\int_{-\infty}^{z}\delta(\theta)d\theta$ is the Heaviside
function.

The solution (\ref{3}) corresponds to Cauchy problem
$W_0(x)=\delta(x-x_*)$. Here we set that $\nu=1$ . The function
$W(x,\tau)$ is in fact a conditional probability distribution, that
is $W(x,\tau)\Delta x\equiv W(x,\tau\mid x_{*},\tau=0)\Delta x$ is
the  probability that at the time  $\tau$ the dynamical variable $x$
belongs to interval $(x,x+\Delta x)$ under condition that at some
previous initial time $\tau=0$     the variable $x$ is equal to
$x_*$.

From the equation (\ref{Ver}) follows that the dynamical variable
has three stationary points:
$$x_{1}=\frac{1}{|p_2|+q_2},\quad
x_{2}=\frac{1}{|p_2|-q_2}, \quad x_3 =0.$$

It is convenient to use the definition $x_1$ and $x_2$ for
transformation of the expression (\ref{3}) to the form:
\begin{equation}\label{4}
\begin{array}{c}
\displaystyle W(x,\tau)=
 \frac{1}{2}e^{-\tau}\left\{\delta\left(x-\frac{e^{\tau}x_*x_2}{x_2+(e^{\tau}-1)x_*}\right) +
 \delta\left(x-\frac{e^{\tau}x_*x_1}{x_1+(e^{\tau}-1)x_*}\right)\right\}+{}\\[2em]
\displaystyle \frac{x_{1}x_{2}}{(x_{2}-x_{1})x^{2}}\left\{ H \left(
 \frac{xx_1 e^{-\tau}}{x_1+x(e^{-\tau}-1)}-x_*\right) -
  H \left( \frac{xx_2 e^{-\tau}}{x_2+x(e^{-\tau}-1)}-x_*\right)\right\}.
\end{array}
 \end{equation}

\section{Calculation of one point moments}\label{sec2}

The one point conditional moments of $n$-th order one defines as
\begin{equation}\label{5}
\kappa_n(\tau)=\langle x^{n}(\tau)|x(0)=x_*,\tau=0\rangle
=\int_{(D)}x^n W(x,\tau)dx,
\end{equation}
where $(D)$ is the support of the probability distribution. Further
we consider the case $(D)=(x_1,x_2)$. After simple calculations one
obtains
\begin{equation}\label{6}
\begin{array}{c}
 \kappa_n(\tau)=
\displaystyle \frac{1}{2}e^{-\tau}\left\{\left(
\frac{e^{\tau}x_*x_2}{x_2+(e^{\tau}-1)x_*}\right)^{n} +
 \left(\frac{e^{\tau}x_*x_1}{x_1+(e^{\tau}-1)x_*}\right)^{n}\right\}+{}
   \\[2em]
\displaystyle \frac{x_{1}x_{2}}{(x_{2}-x_{1})(n-1)}\left\{ \left(
x_2\beta(\tau)\right)^{n-1} -
 \left(x_1\gamma(\tau)\right)^{n-1}\right\},
\end{array}
 \end{equation}
where $$\beta(\tau)=\frac{x_*}{x_* +(x_2-x_*)e^{-\tau}},$$
$$\gamma(\tau)=\frac{x_*}{x_* -(x_*-x_1)e^{-\tau}}.$$
Let $\tau\rightarrow 0$, then $\beta
(\tau)\rightarrow\frac{x_*}{x_2}$ and $\gamma
(\tau)\rightarrow\frac{x_*}{x_1}$. In this limit from (\ref{6}) one
has $\kappa_n(\tau)\rightarrow x_{*}^{n}$.  Let us consider another
asymptotic $\tau\rightarrow\infty$. From (\ref{6}) one obtains for
$n=1$ the following  stationary value of the moment
$$\kappa_1(\tau)=\frac{x_{1}x_{2}}{x_2-x_1}(\ln x_2 -\ln x_1),$$
and for $n\neq 1$ stationary values of moments are equal to
$$ \kappa_n(\tau)=\frac{x_{1}x_{2}}{n-1}
\left(x_{2}^{n-2}+x_{2}^{n-3}x_{1}+...+x_{2}x_{1}^{n-3}+x_{1}^{n-2}\right).$$

Generally the moments $\kappa_n(\tau)$ are given functions depending
on the decreasing exponent $e^{-\tau}$ and can be represented by a
series over the powers of  $e^{-\tau}$. In \cite{Brey} a similar
representation was found for the stochastic Verhulst model with
fluctuating coefficient at the first degree of the dynamical
variable $x$.
 In the limit for large $\tau \gg 1$ the time behavior of $\kappa_n(\tau)$
can be described by a single exponent.

Important role  is played by the first two initial moments
($n=1,2$). Let us consider the moment of first order. In this case
\begin{equation}\label{7}
 \kappa_1(\tau)=\frac{x_*}{2}
 \left(\frac{1}{1+(e^{\tau}-1)\frac{x_*}{x_2}}+\frac{1}{1+(e^{\tau}-1)\frac{x_*}{x_1}}\right)+
\frac{x_{1}x_{2}}{x_{2}-x_{1}}\ln\frac{x_2\beta(\tau)}{x_1\gamma(\tau)}.
\end{equation}
In the asymptotics $\tau\rightarrow\infty$ from (\ref{7}) in first
order over the infinitesimal $\exp(-\tau)$, one has

\begin{equation}\label{8}
 \kappa_1(\tau)\approx\frac{x_{1}x_{2}}{x_2-x_1}\ln\frac{x_2}{x_1}+
 \left(\frac{x_1+x_2}{2}-\frac{x_{1}x_{2}}{x_*}\right)e^{-\tau}.
\end{equation}

It is interesting that there exists  the initial value 
$x_*=\frac{2x_{1}x_{2}}{x_1+x_2}\equiv\frac{1}{|p_2|}$. In this
case the coefficient at $\exp(-\tau)$ is equal to zero. Therefore
we should take into account the next order, i.e. $\exp(-2\tau)$.
 Physically it means that in point $x_*=\frac{1}{|p_2|}$  the
correlation of variable $x(\tau)$ with the given initial value  of
variable $x$ ($x(0)=x_*$) decreases more rapidly  at
$\tau\rightarrow\infty$. When $x_*\neq\frac{1}{|p_2|}$ the
correlations tends to stationary level as $\exp(-\tau)$.

Here we give also an explicit expression for the case $n=2$:
\begin{equation}\label{9}
 \begin{array}{c}
\displaystyle \kappa_2(\tau)=\frac{1}{2}e^{-\tau}\left [ \left
(\frac{e^{\tau}x_{*}x_{2}}{x_{2}+(e^{\tau}-1)x_{*}}\right )^{2}+
\left (\frac{e^{\tau}x_{*}x_{1}}{x_{1}+(e^{\tau}-1)x_{*}}\right
)^{2}\right ]+{}\\[2em]
\displaystyle
\frac{x_{*}^{2}x_{1}x_{2}(1-e^{-\tau})}{(x_{*}+(x_{2}-x_{*})
e^{-\tau})(x_{*}-(x_{*}-x_{1})e^{-\tau})}.
\end{array}
\end{equation}

\section{Concluding remarks }\label{sec3}

We have considered the  time evolution of one point moments of
dynamical variable corresponding to the stochastic Verhulst model.
The explicit form of the moments shows that the moments are the
functions of $\exp(-\tau)$. In \cite{Brey} it was shown for Verhulst
model when parameter fluctuates  at dynamical variable $x$ (not
$x^2$), that the exact solution for one point moments is presented
by a series over powers of $\exp(-\tau)$.  From  formulae obtained
here one can write the moments $\kappa_n(\tau)$ in the same form. It
should be noted that in this communication we have obtained an
explicit form of the solution. Under the condition
$\tau\rightarrow\infty$ the moments decrease in time as single
exponential function.  It is shown that the time dependence of the
moment $\kappa_1(\tau)$ which physically describes the correlation
between the value of the dynamical variable $x$ at the time $\tau$
with its given (nonrandom value $x_*$) at the initial time $\tau=0$
changes. This time behavior depends on the choice of the initial
value $x_*$. The critical value is $x_{*}=1/|p_{2}|$. For this value
of $x_*$ in the limit $\tau\rightarrow\infty$ the correlations
decrease as $\exp(-2\tau)$, not as $\exp(-\tau)$.

It should be noticed that the one-point moments for some special
type of the dynamical system (\ref{Ver}) with polynomial functions
$p(x)$ and $q(x)$ Gaussian white noise fluctuation coefficient at
the first power of $x$ were considered in \cite{BB82,GSch82,GH82},
where it was shown that the asymptotic behavior of the moments is
described by a power function.

\end{document}